\documentclass[conference,10pt]{IEEEtran}  %{article}

\usepackage{array}

\usepackage{algorithm}
\usepackage{algpseudocode}
\usepackage{amsmath}
\usepackage{amsthm}
\usepackage{arydshln}
\usepackage[short,nodayofweek,level,12hr]{datetime}
\usepackage{fancyhdr}
\usepackage{flushend}
\usepackage{graphicx}
\usepackage{marvosym}
\usepackage{semantic}
\usepackage{stmaryrd}
\usepackage{syntax}
\usepackage{textcomp}
\usepackage[matrix,arrow,curve]{xy}
\usepackage{comment}

\begin{document}
\xyoption{all}

\title{Review of Recent Heap Specification and\\ Verification Techniques\\ \small{(1st draft  submitted on 12th March 2015; journaled 2019, self-translated into English)}}

\author{
  \IEEEauthorblockN{\parbox{5cm}{\begin{center}Ren\'{e} Haberland\\Independent researcher\end{center}}\\[-0.3cm]}
  \IEEEauthorblockA{
     \parbox[t]{7cm}{\begin{center}Saint Petersburg Electrotechnical University\\ Prof. Popova Street 5,\\ 197022 Saint Petersburg, Russia\\ haberland1@mail.ru\end{center}}\qquad \qquad
     \parbox[t]{6.5cm}{\begin{center}Carl Zeiss Research and Development\\ Semiconductor Manufacturing Technology\\ 73447 Oberkochen, Germany\\ rene.haberland@zeiss.com\end{center}}
  }
}

\maketitle  %%

\begin{abstract}
The article provides an overview of the existing methods of dynamic memory verification; a comparative analysis is carried out; the applicability for solving problems of control, monitoring, and verification of dynamic memory is evaluated.
This article is divided into eight sections.
%
%%%
The first section introduces formal verification, followed by a section that discusses dynamic memory management problems.
The third section discusses Hoare's calculus resumed by heap transformations to the stack.
The fifth and sixth sections introduce the concept of dynamic memory shape analysis and the rotation of pointers.
%%%
The seventh is on separation logic.
The last section discusses possible areas of further research, particularly the recognition at recording level of various instances of objects; automation of proofs; "hot" code, that is, software code that updates itself when the program runs; expanding intuitiveness, for instance, on proof explanations.
\end{abstract}

\begin{flushleft}
\end{flushleft}

%\thispagestyle{fancy}
%\cfoot{This work is licensed under the Creative Commons Attribution License (CC BY-NC-SA 4.0).}

\IEEEpeerreviewmaketitle  %%

%%%%%%%%%%%%%%%%%%%%%%%%%%%%%%%%%%%%%%%%

\section{Introduction}
For over a couple of decades, mistakes based upon dynamic memory remained crucial during software engineering.
Error localisation is challenging since diagnosed error locations are too often too far from the actual cause.
Errors affect the overall behaviour of an application, its performance and correctness.

This paper overviews and compares existing techniques on dynamic memory verification.
It evaluates applicability for practical verification control.
Before introducing to dynamic memory models, research objectives shall be announced first.

Some running process allocates a \textit{region} of dynamic memory by the operating system \cite{love10}.
The region's memory is allocated on demand (e.g. by \texttt{malloc} or \text{new}) and is denoted as "\textit{heap}" in fig.\ref{FigureMemoryRegions}.

\begin{figure}
\begin{center}
\begin{tabular}{|c|}
 \hline
 \parbox{5cm}{\xymatrix{\textnormal{stack}}}\\
 \hdashline
 \parbox{5cm}{\xymatrix{\textnormal{heap}}}\\
 \hline
 uninitialised data\\
 (\texttt{bss})\\
 \hline
 initialised data\\
 (\texttt{data})\\
 \hline
 program code\\
 (\texttt{text})\\
 \hline
\end{tabular}
\caption{Typical memory regions for a process}
\label{FigureMemoryRegions}
\end{center}
\end{figure}

For clarity, first, a process's memory region shall in the following refer to "\textit{dynamic memory}".
Second, a logical data structure (in the most common case, a graph) being mapped onto a continuous logical memory space refers to "\textit{heap}".

Program code contains processor instructions, namely opcodes with operands.
The initialised data region contains program variables having initial denotational values before that program is run.
Conversely, uninitialised variables are global and non-local variables that are not assigned initial values.
Between stack and heap, there is a floating border.
So, the stack grows when before entering some procedure and shrinks when leaving it.
More accurately, the stack alters, and for control purposes, the stack index pointing to the actual stack frame is of most interest.
Local variables (or locals) are automatically accommodated into the stack frame and freed when leaving.
So, locals are automatically allocated and deallocated whilst program execution.
Unlike stack, the heap is allocated manually and remains unless freed manually.
In 32-bit systems supporting virtual memory are based on segmentation.
It implies that accessory OS control units access memory cells in segments.
In 64-bit OSes segmentation separators still exist, though they are not much used anymore since the need for segmentation steadily diminishes in the case of sufficiently big enough address spaces.

A verification system is a formal checking process, originated from Hoare \cite{hoare69}, which is used in algebraic and logical formulae and logical rules for the proof of (in-)correctness obeying some given specification (so-called "\textit{Hoare calculus}").

Let us consider the following C program

\begin{center}
\begin{verbatim}
list_elem *temp;   list_elem *y = NULL;
while (x != NULL) {
  temp = x->next;
  x->next = y;
  y = x;
  x = temp;
}
\end{verbatim}
\end{center}

which inverts a linked list defined as

\begin{verbatim}
struct list_elem{ 
  int data;
  list_elem *next;
};
\end{verbatim}

After each loop iteration, the dynamic memory and pointers \texttt{x} and \texttt{y} denote as

\begin{center}
\begin{tabular}{l}
  Iteration №1:\\
  \xymatrix{
    \txt{y:} \ar[r] & \texttt{0} \quad
    \txt{x:} \ar[r] & *+[F] \txt{1} \ar[r] & *+[F] \txt{2} \ar[r] & *+[F] \txt{3} \ar[r] & \texttt{0}
  }\\[0.3cm]
  Iteration №2:\\
  \xymatrix{
    \txt{y:} \ar[r] & *+[F] \txt{1} \ar[r] & \texttt{0} \quad
    \txt{x:} \ar[r] & *+[F] \txt{2} \ar[r] & *+[F] \txt{3} \ar[r] & \texttt{0}
  }\\[0.3cm]
  Iteration №3:\\
  \xymatrix{
    \txt{y:} \ar[r] & *+[F] \txt{2} \ar[r] & *+[F] \txt{1} \ar[r] & \texttt{0}  \quad
    \txt{x:} \ar[r] & *+[F] \txt{3} \ar[r] & \texttt{0}
  }\\[0.3cm]
  Iteration №4:\\
  \xymatrix{
    \txt{y:} \ar[r] & *+[F] \txt{3} \ar[r] & *+[F] \txt{2} \ar[r] & *+[F] \txt{1} \ar[r] & \texttt{0} \quad
    \txt{x:} \ar[r] & \texttt{0}
  }
\end{tabular}
\end{center}

When the program starts execution, \texttt{x} contains the initial list, and \texttt{y} is empty.
When finalising the loop, \texttt{y} contains the list in reverse order, and \texttt{x} becomes empty.
Here we notice that the semantics of this pretty simple program is far from trivial, in any case.
A specification's main objective is to describe the dynamic memory before and after a program runs.

\section{Problem Actuality}
\label{sect:DifferenceStackAndHeap}

The main difference between stack and heap is the storage of data.
The program manages stack (indirectly, but) automatically at two locations: when entering and when leaving a procedure.
Heap is managed manually only using special operators.
The memory content alters permanently on each execution step.
Heap can be denoted by an oriented graph whose vertices are structures containing application-specific data and whose edges represent pointers.
Fig.\ref{FigureHeapOrganisation} demonstrates a typical heap, where fig.\ref{FigureStackOrganisation} shows a stack containing the stack-local variable $o$.

\begin{figure}
\begin{minipage}[b]{0.3\linewidth}
\centering
\begin{tabular}{c}
\xymatrix{
 o \ar[r] & *++[Fo]\txt{'b'} \ar[dr] & o2 \ar[d]\\
 *++[Fo]\txt{333} \ar[rr] \ar[ur] & & *++[Fo]\txt{D}
}
\end{tabular}
\caption{Heap $o$,$o2$}
\label{FigureHeapOrganisation}
\end{minipage}
\qquad \qquad \qquad
\begin{minipage}[b]{0.3\linewidth}
\centering
\begin{tabular}{l}
\xymatrix @W=3pc @H=1pc @R=0pc @*[F-]{
1 \save+<-4pc,1pc>*\hbox{\it o}
\ar[]
\restore \\
{\bullet}
\save*{}
\ar`r[dd]+/r4pc/`[dd][dd]
\restore \\
{\bullet}
\save*{}
\ar`r[d]+/r3pc/`[d]+/d2pc/
`[uu]+/l3pc/`[uu][uu]
\restore \\
456 }
\end{tabular}
\caption{Stacked local}
\label{FigureStackOrganisation}
\end{minipage}

\end{figure}

Heaps are built up stepwise.
The size of a heap vertex is determined by its corresponding data type.
Incremental built-up implies alterations may remain unnoticed even if pointers are untouched.
In C, heap vertices are managed by the program statements \texttt{malloc} and \texttt{free} (cf. \cite{weihl80}).
Pointers accommodated in dynamic memory have access to heap vertices and are syntactically described by heap expressions.
Unreachable heap elements are considered garbage and shall be dismissed.
Pointers sharing at least a typical heap cell are called "\textit{aliases}".

Unsurprisingly, working with dynamic memory imposes many practical issues despite its initial vague complexity.
Those were firstly classified by Miller \cite{miller95}, and they remain up to date.
The paper \cite{miller90} titles dynamic memory problems as one of the most challenging issues during programming and software engineering.
It contains an illustrative comparison of software development tools under Linux and Windows.
Essentially, it was found that Linux software components contain fewer errors than corresponding components being implemented in Windows.
The success can be explained by the active support of the Linux community.
Miller announces that both invalid addressing and invalid boundaries are the most frequent error related to dynamic memory.
However, frequently problems may not easily be located and may be altered at different locations accidentally.
For instance, if invalid heap access causes an immediate crash, this is ideal from a diagnostic standpoint.
In other cases, however, localising errors due to invalid pointer operations may become quite challenging.
One approach in reducing the complexity of the dynamic memory scope is transforming a heap into a stack (cf. sec.\ref{sect:HeapIntoStack}).

Heap problems may be categorised by their features, for instance, as follows:
\begin{itemize}
	\item [(i)] memory leakage;
	\item [(ii)] inaccessible heap; 
	\item [(iii)] invalid heap access;
	\item [(iv)] deviation between given and expected data structures; 
	\item [(v)] aliasing and "\textit{remote alteration}".
\end{itemize}

The following sections go into more depth on each of the mentioned problem.

\subsection{Memory leakage}

Let us consider the example
\begin{verbatim}
MyClass object1=new MyClass();
...
object1=new MyClass();
\end{verbatim}

Let us further assume there are no pointers that would point to the previously allocated memory region between the first and second assignment of \texttt{object1}.
According to Weihl \cite{weihl80}, "\textit{garbage}" is defined as a heap that is not referenced by any pointer.
The first object of class \texttt{MyClass} is lost --- essentially, it just becomes garbage.
If the object is no more in need, it shall be recycled.
Otherwise, repeatedly allocated regions may cause the applications and OS to fail.
It may either stop execution prematurely and just quit with some error or continue and slow down the overall OS due to a search for free memory resources.
Usually, the latter is hard to debug in practice due to its non-deterministic behaviour.

Jones et al. \cite{jones11} give a concise overview of garbage collection and an in-depth discussion of recent techniques focused on multi-threaded garbage collection.
Furthermore, estimates are made on the most promising methods.

Appel \cite{appel87} is most concerned about von-Neumann's computer architecture regarding pushing and popping stack frames (cf. sec.\ref{sect:HeapIntoStack}).
He is confident that dynamic memory can have significant performance advantages over stack (cf. \cite{tofte97}, \cite{meyer2-03}).
Appel's multi-threaded approach utilises a "copying garbage collector" \cite{jones11}.
%.
The collector is triggered only when data in scope is altered, and the total amount of allocated data is multiple times greater than the remaining amount of free heap.
Regardless of the age of the article \cite{larson77}, the distinction between fast cache and slow but more memory remains up-to-date.
Larson considers a "condensed collector" operating in dependency of the number of disposable regions ($R$), active regions ($A$), and size of fast memory ($H$).
He postulates two optimal strategies (for performance) for allocation and deallocation of dynamic memory:
\begin{itemize}
\item  maximisation of $R$, if A$\ll$H does not hold;
\item  assigning R:=H, if A$\ll$H. 
\end{itemize}

Apart from the restrictions in \cite{appel87}, there is one more related to addressing:
"XOR"-linked data structures (see \cite{sinha04}, \cite{parlante01}) do not allow efficiently locating garbage as described in \cite{jones11}.
It is mainly because heap regions (e.g. structures, pointers to linked list elements) addresses are no more absolute; they are relative.

Assume --- that $A_{XOR}$ denotes the finite linear address space, then $(A_{XOR},\oplus)$ denotes the group with well-known rules for associativity, neutral and inverse elements, and enclosed operation $\oplus$ \cite{haberland16-6}.
Any specified address of a dynamic memory may be calculated directly and inversely.

For instance, given $a$ and $b$, then $a \oplus (a \oplus b) \equiv b$, and $(a \oplus (a \oplus b)) \oplus (a \oplus b) \equiv b \oplus (a \oplus b) \equiv a$.
By applying XOR-expressions, the second pointer may be skipped by an auxiliary arithmetic operation  (cf. \cite{sinha04}, \cite{parlante01}).
This saving may be helpful in collectors, especially in an embedded background with a limited amount of memory.
However, \cite{sinha04} contains mistakes, and its calculation model presented is incomplete.
Fortunately, a corrected variant may be found in \cite{haberland16-6}.

Meyer \cite{meyer2-03} suggests a permanently enduring garbage collection.
This installation can hardly be implemented in a single-threaded environment due to a severe lack of resources causing unacceptable delays.
Undoubtedly a better solution would be to localise the reasons for garbage generation and rewrite the critical code section instead to avoid garbage in the first place.
In that case, efforts in the collection would nullify.
In multi-threaded environments \cite{jones11} impressively shows, may work more efficient if the collector is turned on periodically and if a realistic heuristic of estimated garbage determines that frequency.
Otherwise, garbage collection may not redeem estimates even close.
However, even in embedded environments, memory steadily increases, so an optimal collector becomes less and less critical after all by practical means, except leaks that endanger the overall system stability.
Hence, detecting "\textit{hot locations}" is worth the effort --- from a practical point of view, this is code enclosed in loops, which is often executed.

The paper \cite{sun06} suggests garbage collection \textit{by generation}.
Generations depend on the frequency and duration certain heap entities are accommodated.
If some object resides for an extended period and is referenced frequently, it should be relocated to a quicker memory region.
The authors' comprehensive research shows their chosen performance heuristics approximate to an optimum.
Garbage collection is used by default in programming languages like Java, C\# and in many popular functional programming languages and is turned off in ISO C and C++ \cite{isocpp14}.

\subsection{Invalid access / inaccessible memory}

Incorrect memory access may occur whenever a procedure is granted full access by pointer, even when the alteration may take place later (unintentionally).
Let us consider the following example

\begin{verbatim}
// object1 not constructed here
MyClass object2=new MyClass();
object2.ref=object1;
// object1 constructed
\end{verbatim}

There is a reference to \texttt{object1}, although it is not yet constructed.
This error is not detected during the second assignment but the first use of \texttt{object2.ref}.

\subsection{Incorrect memory access}

In the given examples, we are interested in invalid heap access.
Let us assume an object reference to \texttt{object1} is initialised and field \texttt{ref} equals \text{NULL} whilst running.

\begin{verbatim}
// object1.ref==null
value = (object1.ref).attribute1;
\end{verbatim}

Then, in the best case, the program fails, indicating incorrect access to the segment \texttt{heap}.
During verification the input program is checked for error until execution.
If the given field is not always initialised, this may imply uninitialised objects, which must be avoided.

Also, incorrect access occurs due to incorrect heap addressing, as it may occur with arbitrary addressing.
If access is granted strictly according to existing class fields, it only needs to be checked whether access always exists to dynamic memory during runtime.
If pointer access depends on an arbitrary arithmetic expression being calculated on runtime, then checking dynamic memory becomes undecidable.
The most common case is rather uncomfortable.
However, when restricting to apriori known class fields only, expressibility does not suffer much (cf. sec.\cite{bornat00}).
All heap regions need to be monitored to avoid incorrect access.

Let us now consider the following example.

\subsection{Deviation from a data structure}

In the following example, some object is printed to the console.

\begin{verbatim}
object1.next=object1;
...
root=object1;
while(root.next!=null){
  printf(%d, object.data);
  root=root.next;
}
\end{verbatim}

If a simply-linked list with a specified start and end are given, then the program can inductively be found to be correct.
However, if the given data structure contains a cycle unexpectedly the calculus may become incorrect according to the specification provided.

Recommended references on object theory may be found in \cite{cardelli96}, \cite{gunter94}.
There, an axiomatic formal base for object classes is founded.
According to Abadi and Cardelli, objects are instantiated classes, which may have different interpretations for subtypes of a class.

In \cite{abadi97}, objects are considered not as abstract data types but as simple records as defined by \cite{ehrig80}.
Abadi does not introduce recursively-defined classes and skips pointers nor object pointers (nor aliases).
In the model, proposed objects are non-sharing heap regions.
Class types ($T$) are defined recursively.
Those are either integer or are compound.
Compound types are classes themselves that may be compound again.
A class object contains a set of unique fields and methods.
Fields are of type $T$, and methods are of type $T_j \rightarrow T_{j+1} \rightarrow \cdots \rightarrow T_k$.
The check if an object is an instance of a class or subclass is defined element-wise of all class fields and methods.
A unique result register can describe the states before and after a program statement.
Recursively defined predicates are disallowed by \cite{abadi93} since there are rigid theoretical boundaries for that data model.
In \cite{leino98} a relaxation for that boundary is proposed defined over an algebraic ideal ring construction.
However, this construction may lead to incorrect verifications due to multiple derivations for the recursive cases due to the object combinator definition.
This onerous restriction holds for any practical.

Moreover, this model does not separate the heap and lacks fundamental incompleteness (cf. sec.\ref{sect:HoareCalculi}).
Banerjee \cite{banerjee08} proposes a language that describes stacked objects only, and those objects must be of one type (cf. \cite{regionmem10}, \cite{tofte97}).
The approach in \cite{banerjee08} (see sec.\ref{sect:HeapIntoStack}) moves all locals into the stack since pointers are forbidden.
Recursively-defined predicates over objects are forbidden.
Global invariants are unique since they catch the fixed dependencies between objects.
Banerjee warns about the increasing abstraction problem and supports the initiative of predicates describing different aspects of an object.

Both \cite{barnett04} and \cite{mueller02} categorise underpinning the object-oriented modelling "\textit{Unified Modeling Language} (UML)" \cite{dsouza98}.
Apart from the graphical modelling, "\textit{Object Constraint Language} (OCL)" \cite{oclspec} represents a textual record.
Formulae in OCL describe class objects and their dependencies.
The given expressibility can be equivalent to the second-order typed $\lambda$-calculus (as introduced by the "\textit{System F}" verifier).
However, "UML" and "OCL", which are modelling languages, cannot express aliases or pointers to objects in heaps --- which is a significant limitation.

\subsection{Remote object alteration and aliases}

The following example demonstrates two aspects.
First, \texttt{const} does not necessarily protect the content of a heap cell from tampering, even if protecting it with an additional \texttt{const} keyword as shown for pointer \texttt{pa2} in the following:

\begin{verbatim}
int *x;
int *good() { int *p = x ; return p; }
int *bad()  { int x = 55; return &x; }

void main(){
  int a=5, b=4;
  int *pa = &a;
  const int *const pa2 = &a;
  a = 77;
  pa2 = pa;
/***  ’ pa2 ’ is read - only ***/
  printf(" *pa:%d, *pa2:%d \n" ,*pa, *pa2);
/***  *pa==77, *pa2==77    ***/
  a = *&b;
/***  *const of pa2 is still no 
      guarantee for safety   ***/
  printf(" *pa:%d, *pa2:%d\n", *pa, *pa2);
/***  ! problem: *pa==4, *pa2==4   ***/
  x = (int*)malloc(10);
  memset(x,7,sizeof(x));
  printf("0x%x\n", *(good())); // OK
  printf("0x%x\n", *(bad())); // SEGV
}
\end{verbatim}

Second, the life span of variables and their values in a heap dramatically differs from stacked variables.
So, assigning locals may be correct locally.
However, when trying to access a previously freed stacked local address may lead to an application halt depending on whether stack protection is granted by the OS (refer to both functions \texttt{good} and \texttt{bad} in the example).

When two pointers point to a shared heap cell, both pointers are aliases to each other.
However, the fast and reliable detection of aliases remains an actual and open problem.
When a procedure is called, there is always a "callee" (inner) and a "caller" (exterior) side.
For instance, algorithms for inner analyses can efficiently be constructed with less effort, Muchnick's algorithm family \cite{muchnick07}.
However, exterior analyses may, in general, be very complex because the whole transitive hull of calls needs to be analysed, which grows exponentially per caller level \cite{ramalingam94}.
Another tricky question relates to the decision of whether a pointer strictly aliases or not --- where a general method may become imprecise, ineffective or both.
In practice, the "GCC" and "LLVM Clang" both implement auxiliary compilation switches to resolve code optimisations, for instance, by facilitating the \texttt{-fstrictaliasing} switch.
The programmer may provide the keyword \texttt{__restrict__} to indicate that passed pointers may not be treated as unbound pointers.
Although this is precisely the desired behaviour in over 95\% of all cases, the remaining tiny percentage still may generally cause devastation if not treated correctly, which a compiler must by default take into consideration, causing worse code to be generated than could be with that keyword.
The given keyword may be used with objects on a procedural level too which heavily may improve performance since the objects usually will be flushed once leaving a procedure and do not require further synchronisation between procedures \cite{gcc15}.\\

Weihl \cite{weihl80} gives a broad introduction to aliases regardless of the article's age (see later on).
According to Weihl, alias analysis approximates a pointer's scope spanning an exponential search space of locations that may tamper its content.
So, beyond a procedure's body, the analysis when a pointer's content may be altered becomes much more complicated.
Here, the locations and the call itself need to be carefully analysed in a pointer's continuation.
Naturally, continuation evaluations are tractable.
Weihl suggests a metric for alias analysis tractability.
It shall be noted here that metric ideally should also be used to feedback a developer to schedule a reengineering task.

Muchnick \cite{muchnick07} provides a succinct overview of existing and previous alias analyses, including his methods.
He divides the analyses into control graph dependent and independent techniques.
He further divides the analyses by its outcomes, namely whether a pointer "strictly is alias", "maybe alias", and "strictly no alias".

Horwitz \cite{horwitz89} boosts furthers Muchnick's alias analysis by introducing bit-vectors to encode the states "\textit{alias/may alias}" and "\textit{strictly no alias}".
Furthermore, generalised analyses are found in \cite{khedker09}.
Here, Khedker also uses bit-vectors for his most generalised control-flow based analyses.
Khedker's approaches are based on the well-known Floyd-Warshall algorithm for checking reachability in directed graphs (cf. \cite{cormen09}).
Moreover, the author of this article suggests Khedker's approach shall be further extended to incorporate "SSA"-forms (cf. \cite{cytron91},\cite{ssabook15}) in the most general case.
Naeem \cite{naeem09} seems to confirm my personal opinion in a minor remark.
Naeem further suggests incorporating hash-tables in order to improve bit-vector operations.
Here, it must be noted that Naeem's proposition is a desire to redefine alias analysis as such as a problem of "SSA".
The author of this paper strongly admits his proposition.
At least, most definitely, further research is required.
It would almost certainly trim a whole phase of static analysis when found tractable, causing heavy performance gains due to a full and more effective register referencing.
So, notoriously redundant copying from minor til big memory heap chunks may be dropped that are still required for security.

Pavlu \cite{pavlu10} divides alias analyses into two categories: approaches based on "unification" \cite{steensgaard96}, which are more accurate in finding aliases, and approaches based on "approximation", which are less accurate but much more straightforward.
The latter approaches can be considered a compromise rather than a fully qualified exact solution.
Whether an accurate or an approximate solution is efficient remains open after all.
In other words, it would be interesting to find a scenario where aliases are not wholly excluded, but data structures may be altered, such that the modified memory chunk is caught inside a procedure.
The author of this paper believes this question has not yet been entirely covered.

\section{Hoare Calculus}
\label{sect:HoareCalculi}
Hoare first introduced a formal calculus in 1969 related to specification and software verification \cite{hoare69}.
It is fundamental up-to-date, and its basic theoretical approach has not changed ever since, except extended multiple times.
An axiomatic semantics and operational semantics are proposed as a toolbox for describing programming languages.
Hoare's axiomatic semantics denotes which rules are applied for an input programming language and which rules need to be applied to prove its correctness.
It compares its intermediate denotations with those of rules.
If both state denotations match, a given program obeys a given specification.
Otherwise, the program does not obey the rules for correctness.
Hoare includes algebraic and logical assertions in expressions.
In honour of Hoare, $P\{C\}Q$ is called "Hoare's Triple", which nowadays is often rewritten to $\{P\}C\{Q\}$, where $P$ denotes the precondition, $C$ a program statement, and $Q$ denotes the postcondition.

Assume, $\Gamma$ denotes the set of logical rules of kind $\frac{A}{B}$, where $A$ denotes the antecedent, and $B$ the consequent.
An axiom is a rule whose antecedent is a tautology in the considered area of discussion.
It can be noted as $\frac{}{B}$, or as $\texttt{true} \rightarrow B$.
If a Hoare triple is derivable syntactically from $\Gamma$, then it is written as $\Gamma \vdash \{P\}C\{Q\}$ or as $\vdash_{\Gamma} \{P\}C\{Q\}$.
If clear from its context, only $\Gamma$ is considered, then sometimes $\vdash_{\Gamma}$ can be dropped for simplicity.
A Hoare triple is interpreted as a predicate.
This predicate is only determined when $C$ terminates (or --- is a terminal in a more generalised way).
The interpretation is proper whenever the precondition $P$ is true, statement $C$ is run, and postcondition $Q$ succeeds.
Otherwise, the Hoare triple is not correct.
In case $C$ does not terminate, the Hoare triple is undefined.
A Hoare triple interprets as false whenever a postcondition fails.
The designers and test engineers define conditions.
If $Q$ is incorrect for a given program, any result obtained shall be doubtful and require further clarification.
$P$ and $Q$ are formulae describing the calculation state at any time.
So, there must be references to symbols in the calculation state, variables and symbols in $C$.
Although Hoare did not assert it himself, it seems highly plausible to use some formal logic to express and reason rules about a given program.

In 1969 Hoare noted that some program statements in $C$ are much harder to describe than others and, as a direct effect, much more demanding to prove its correctness.
Such statements encompass labels, unconditional jumps, and unbound parameters.
Still, today this problem remains essential, and further features emerge even more.
Due to software modularisation, labels are often excluded in high-level programming languages nowadays from verifications entirely.
Apt \cite{apt81} and Clarke \cite{clarke79} analysed unbound parameters in detail and found that this problem nearly vanished today in Algol-styled programming languages.

In general, proof continues until all branches succeed with axioms or until proof can be refuted.
The overall structure of proof is a tree.
Whenever a cycle occurs in a proof, the proof fails.
If, however, previous subproofs are utilised, copy subproofs can be skipped.

A proof starts with a consequent and searches for matching conditions in antecedents for each assertion until entirely derived to axioms.
Here, $C$ is broken down until the remaining program statement is empty or contains only basic statements.
It is worth noting that the programming language is kept open by Hoare; often, it is imperative, however.
Here intermediate calculation steps (the memory states a program is at a particular line of code) are transformed stepwise per statement.
So, there is no reason why a descriptive programming language may be used instead.
Apart from very dedicated functions (e.g. system calls), there are also theoretic boundaries imposed to verification rules.
One such problem is the guessing of the right loop invariant.
Let us consider a rule for loops:

\begin{center}
\begin{tabular}{c}
  \inference{\{p \wedge e\}S\{p\}}{\{p\}\texttt{while }e\texttt{ do }S\texttt{ od}\{p\wedge \neg e\}}
\end{tabular}
\end{center}

Here, $e$ denotes a condition, $S$ is a program statement, and $p$ denotes the precondition.   %% TODO: !check again paragraph !!!
From the consequent follow, the initial condition $e$ must be false when leaving the loop.
However, all conditions unifying in $p$ before the loop must be valid and after the loop is executed.
In order to prove the correctness of that loop, $p$ with the correct loop body and initial condition $e$ shall be shown because the loop's body is accessed only as long as $e$ holds.
After $S$ $e$ may be true or false.

In order to fully specify a loop, all variables occurring in its body would be required.
Once a loop is executed, the calculation state alters, and naturally, a subset of all variables need to be adapted to describe the current state.
Formulae describing calculation states after an arbitrary number of loop iterations manifest the invariant.
Loop invariants due to theoretical boundaries cannot always be fully specified whilst analysing, e.g. in multi-threaded environments.

A minimisation fashion can be observed in many programming languages chosen for verification:

\begin{tabular}{lcl}
 $x:=t$ && .. assignment\\
 $S_1;S_2$ && .. statement sequence\\
 $\texttt{if } e\texttt{ then } S_1 \texttt{ fi}$ && .. conditional jump\\
 $\texttt{while }e\texttt{ do }S\texttt{od}$ && .. loop
\end{tabular}

Strictly speaking, conditional jumps can be replaced by loops too.
From a theoretical point of view, programs composed as such are minimal.
However, expressibility does not suffice \cite{clarke79}.

Apt \cite{apt81} researches Hoare calculi over the last decades and indirectly confirms the trend towards minimal programs as just suggested.
Apt shows in \cite{apt81} and \cite{apt93} why some features have some "undesired" impact towards expressibility and completeness.
It occurs, e.g. due to arbitrary unbound recursion, which can alter the calculation state entirely.
It is also due to recursively-defined data structures that are partially calculated when accessing fields and due to parameters being allowed as parameters in procedure calls.
Those negative impacts have not been resolved yet.

Cook \cite{cook78}, therefore, pleas for a limited subset of Hoare's and Apt's calculi \cite{apt81} and suggests excluding unbound recursion, co-routines, procedures-as-parameters, as well as other non-Algol styled language features (such as lazy data structures).
He proves the following ruleset is complete (according to Cook) and correct due to a constructed abstract machine based on operational semantics:

\begin{center}
\begin{tabular}{c}
 \inference[VARDECL]{P[y/x] \{ \texttt{begin}\ D^{*};\ A^{*}\ \texttt{end} \} \ Q[y/x]}{P\{ \texttt{begin new } x; D^{*}; A^{*} \texttt{end} \} Q}\\\\
 \inference[PROCDECL]{D,\ P\{ \texttt{begin}\ D^{*};\ A^{*}\ \texttt{end}\}Q}{ P\{ \texttt{begin}\ D;D^{*};A^{*} \ \texttt{end} \} Q}\\\\
 \inference[COMP1]{ P \{ A \} Q,\ Q \{ \texttt{begin} \ A^{*}\ \texttt{end} \} R }{ P \{ \texttt{begin} \ A;A^{*} \ \texttt{end} \} R}\\\\
 \inference[COMP2]{}{P\{begin\ end\}P}\\\\
 \inference[ASN]{}{P[e/x]\{x:=e\}\ P}\\\\
 \inference[COND]{P\ \wedge R\{A_1\}Q,\  P\ \wedge\ \neg R\{A_2\}Q}{P\{if\ R\ then\ A_1\ else\ A_2\}\ Q}\\\\
 \inference[WHILE]{P\ \wedge\ Q\ \{A\}\ P}{P\{while\ Q\ do\ A\}\ P\ \wedge\ \neg Q}\\\\
\end{tabular}
\end{center}

\begin{center}
\begin{tabular}{c}
 \inference[CALL]{p(x:v)\ proc\ K,P\{K\}Q}{P\{call\ p(x:v)\}Q}\\\\
 \inference[VSUB]{P\{call\ p(u:e)\}Q \qquad \sigma = z'/z}{P\sigma\ \{call\ p(u:e)\}\ Q\sigma}\\\\
 \inference[CONSEQ]{P > R,\ R\{A\}S,\ S > Q}{P\{A\}Q}\\\\
 \inference[PSUB]{P\{call\ p(x:v')\}Q}{P\ u,e/x',v'\ \{call\ p(u:e)\}\ Q\ u,e/x',v'}
\end{tabular}
\end{center}

Cook intervenes the most practical limitations are: (1) non-termination cannot be taken out inside a concrete Hoare calculus in general, (2) expressibility of the assertion language (for specification and verification) is one of the fundamental problems still open.
It can also be deduced that proofs require simplification since a simple proof is a suitable proof --- this is essential to proof program properties.
Imagine a program was not obeying a given specification, then a simple and generic mechanism is crucial in finding out the reasons for that mismatch.
Of course, parameter substitution suffers disadvantages, such as a clumsy selection at a loop, but this is not the central question for the selected ruleset.
Also important is whether, for a given ruleset and input program, all rule applications taken out in different orderings will lead to the same result or may not get stuck.
This question is about "\textit{proof confluency}" and remains essential and still unsolved up to date.\\

Clarke \cite{clarke79} considers the problems mentioned by Hoare and Apt the single most important ones still valid today.
He lists problems not resolved yet or not resolvable in general in classical Hoare calculi.
For instance, continuations and variable modes (e.g. static, automatic, dynamic and global) are still not sufficiently resolved yet.
On the one side, these problems appear to be quite old now when judging their first appearances in journals.
However, after studying more recent work, one finds they have not been resolved fundamentally yet.
For example, static variables can hardly be described in a Hoare calculus because that would require an efficient notational apparatus to push and pop to the stack, which is not apriori part of the Hoare calculus.
Clarke's completeness definition is not that much different than Cook's.
Both authors understand that some formula is complete if and only if each valid formula is provable.
In addition to Apt \cite{apt81}, Clarke specifies "bad" properties a Hoare calculus may have damaging completeness.
Clarke considers self-application in recursive definitions as a possible reason for damage to correctness and completeness.
Let us recapitulate that any formal system is incomplete, according to G\"odel, since there is always a formalised statement that cannot be proven in terms of that same formal system itself.
Naturally, this affects Hoare calculi too, particularly those related to dynamic memory provers.
Corner cases are essential from a theoretical standpoint, although, from a practical standpoint, those may be incredibly useless since, in most cases, verification still may be done.
It does not affect the fundamental decidability of terms on verifications.
They are bound to addition and subtraction (so-called Presburger arithmetics \cite{presburger29}) but do not allow multiplication or any other operation.
However, this theoretical decidability is insufficient because of the exponential complexity imposed for the worst case.
As a result, practical implementation and theoretical boundaries may diverge quite considerably.
Theoretical bounds are almost neglectable for practical implementations on a day-by-day basis.

Clarke researched a combination of program features that allow or disallow completeness.
For instance, for a given imperative programming language:

\begin{itemize}
 \item[(i)] procedures as parameters,
 \item[(ii)] arbitrary (($\mu$-recursive) procedures,
 \item[(iii)] static variables,
 \item[(iv)] global variables,
 \item[(v)] nested (so-called "inner") procedures,
\end{itemize}

Completeness may be violated in the most general case due to feature iii).
The resulting Hoare calculus is complete if problematic features are excluded (here i-ii, iv-v).
To prove this and similar cases based on the features i-v, Clarke introduces an operational semantics over relations very close to Cook's semantics.
Beyond Clarke’s article, the observations made may be beneficial, for example, on logical predicates and abstract predicates, as shown later in this paper.
Clarke shows that excluding of self-applicable procedure parameters may indeed reduce its expressibility, but it also leads to calculi without sudden incorrect jump statements.
Furthermore, co-routines in combination with arbitrary recursion, however, may lead to incompleteness and incorrectness at the same time.

Meyer \cite{meyer1-03} notices that variables allocated at arbitrary locations in program code dramatically overcomplicate the specification and verification of heaps.
Their elimination may not be considered a solution, though, because any restriction may significantly reduce expressibility.
Moreover, Meyer states that his assignment $a:=b$ is always correct in introducing his Hoare calculus.
However, when looking carefully, one may notice its correctness damages when $b$ contains a procedure call that alters the calculation state.
Hence, it is better not to consider Apt's, Clarke's, and Cook's approaches in isolation and for a rule solely, but to consider always the whole ruleset as mentioned in \cite{clarke79}.
So, the set union $\bigcup$ does not hold in general regarding correctness.

All authors mentioned in this section agree that variables in dynamic memory are a systematical damaging criterion upon all calculi considered yet.
It implies that Miller \cite{miller90}, from his practical investigations and Clarke and all other authors mentioned, independently confirm that dynamic variables are tough to deal with naturally.
Clarke and all other previously mentioned authors still insist on further research to be taken out on dynamic variables.
All warn though this is to be problematic in Hoare calculi.

\section{Transforming into Stack}
\label{sect:HeapIntoStack}

The method "transforming into stack" is a prominent technique \cite{meyer1-03}, \cite{meyer2-03}, \cite{jones75}, \cite{jones11}, \cite{appel87}.
The motivation behind this is that pointers are too often too hard to analyse, and garbage collection requires resources that otherwise would not be needed (cf. sec.\ref{sect:DifferenceStackAndHeap}).
However, Appel \cite{appel87} demonstrates that the latter concern may vanish in cases the algorithm based upon heap is rewritten.
Scenarios are shown where garbage collection is faster than automated stack management.
There is no doubt that algorithm analysis over pointers may be arduous.
However, often algorithms may be rewritten, not necessarily simpler but faster than analogous stack implementations.
So the question may arise: why is then performance important?
In case when passed objects are not touched, copying memory regions consumes a quite considerable overall runtime due to the von-Neumann architecture.
However, even when copies are touched (which often occurs in practice too), "\textit{Application Binary Interfaces}" (ABI) still forbid individual copies to be propagated any further among callee-caller chains.
Currently, simple cases are already recognised by GCC 6.3.0 (cf. \cite{gcc15}).
So, effective use is not possible apriori.
GCC divides objects and structures into smaller chunks, namely single processor words.
It happens mainly without considering class encapsulation as long as data dependencies are concerned, which is not a problem to code generation.
However, if an object ought to be deleted, unfortunately, the whole object remains accommodated for conservative concerns and is not freed even if it could.
The consequence is a bloated execution model.
Mainly, linked objects raise the pressure to free parts more often due to their relative bigger size and remain at the end for security reasons --- which both reduce the chance for improved and minimalistic code significantly.
If an update is applied to an object web, then an object may not be freed. Instead, it shall be passed to the corresponding level.
This problem could be resolved if objects were pushed or pulled to/from the stack simultaneously.
However, this does not happen, and in the end, redundant copies of objects take place everywhere for a conservative approach.

Transforming a heap means any data structure from the heap shall be accommodated to a stack window instead.
Object dependencies shall be codified in the stack by additional constraints.
For instance, all consecutive elements shall be placed with arising indices for a simply-linked list.
The same may be done with doubly-linked lists.
If an element is inserted or removed from a specific position in a list, then the stack window accomodating that list may dramatically differ afterwards to guarantee arising indices properly.
In analogy to that, trees and graphs require "proper stackinisation".\\

Meyer suggests storing all objects in the stack.
Appel and others show this is, in fact, a loss of performance.
Rewriting a heap algorithm into a stack-based algorithm shows poorer runtime characteristics for widely popular data structures.
For example, stack allocation using \texttt{calloc} instead of the heap-specific \texttt{malloc} may reserve several kilobytes on the stack.
It must be informed that this may significantly bloat stack windows; as a result, pages managed by an OS need to be spread more often, which negatively influences performance (cf. "variadic functions" \cite{isocpp14}).

Whether a successful transformation into a stack of objects is not possible to determine apriori.
Moreover, the opposite may even be the case.
Dynamic data structures are preferred over stack whenever the final runtime sizes are unknown.
Apart from that, unrestricted heap manipulations cannot always be transferred as is to the stack since this may require additional constraints modelled differently.
Slower runtime performance, more extensive remaining resources, or rigid implementations are examples.
That is why Meyer's proposition shall be considered with high reservations only.
For completeness, it must be noted that naturally, both programs operating on stack or heap can be implemented in more or less efficient ways, and a direct comparison may not always be valid.
According to Meyer, pointer and alias analysis are complex because the underlying logic does not distinguish between structures and non-structured data in logical expressions.
This equality implies considerable efforts shall be spent in the area of discussion.

A particular case of "Transforming into Stack" may be considered the so-called "Region Calculus" \cite{tofte97}.
The main idea is to assign class objects onto heap as it were locals in a stack window, such that those may be managed automatically.
Since the class object's type is known statically, the corresponding stack windows size may be precalculated successfully.
Even subclass objects may be approximated based on a superclass.
It is worth mentioning that functional programming languages, such as "ML", internally perform precisely this type of coercion.

Nevertheless, regarding the functional approach, one limitation must be kept in sight: no lists may be returned, and the visible object scope must be modelled separately --- since symbols are not bound to local blocks as in classic imperative programming languages.
Another weakness is that regions accommodated in the stack are often too large and are coupled with other objects (where loosely coupled by design seldomly occurs, if ever).
Those categories dramatically slow down performance and the whole verification process.

\section{Shape Analysis}
The main goal of Shape Analysis \cite{sagiv02}, \cite{nielson99}, \cite{pavlu10}, \cite{ramalingam94} is the detection of heap invariants (shapes), which may further be used for alias analysis.
The dependency graph of shapes is described fully by using transfer functions, such as: "shape is empty", assigning a field member or pointer, and allocating a new heap.
Sagiv \cite{sagiv02} suggests a categorisation of pointer relations, namely "aliasing", "non-aliasing", and "possibly aliasing".
Labelling becomes a concern in compressed dependency subgraphs, as discussed in \cite{sagiv02} and \cite{nielson99}.
The approaches discussed in there have numerous limitations, e.g. pointers may only be indexed in non-variable expressions.
Furthermore, pointers to class objects are forbidden.
Arrays with a dynamic length are also forbidden and sharing structured (as "\texttt{union}" does in C).

Pavlu \cite{pavlu10} notes that both approaches, \cite{sagiv02} and \cite{nielson99}, may conclude in unsound reasoning for a given program due to non-deterministic choices to be made on approximation.
If for "if-then"-statements in one case "possibly aliasing" is found, but in an alternative case "aliasing", then according to their method, the calculation would terminate with founding "aliasing", although the right should instead be "possibly aliasing".
Apart from that, \cite{pavlu10} contains a detailed comparison of \cite{sagiv02} and \cite{nielson99}.
Pavlu judges the method in \cite{nielson99} as most accurate.
In addition to it, he suggests a path optimisation for shape graphs for the same beginnings and aliases.
The upper complexity bound for his proposition is quadratic in dependency of the incoming path length and reduces the overall calculation by approximately 90\%.
Pavlu suggests a complex alias analysis beyond procedure definitions by turning procedure calls and globals as much as possible into locals.
This approach is not new, and it can be found that it has already successfully been applied to "\textit{GCC}".
Pavlu and the author of this paper suspect context-independent approaches may lack a future perspective due to imprecision (cf. also \cite{hind01}).
Furthermore, Pavlu suggests an additional optimisation of unifying vertices, generating subgraphs based on vertices that do not contain aliases.

Parduhn \cite{parduhn08} proposes an environment for visualising heap shapes for faster invariant analysis and close-invariant relationships (those shape relations that seldomly correlate).
In order to navigate through that visualised shapes, two basic operations are used: "shape abstraction" and "shape concretisation", which allow for zooming out and in subgraphs.
However, the visualisation of the transfer function that allows zooming in/out more than one shape currently remains unsatisfactory, and there is no indication this could generally improve in the next future due to its nature.

Calcagno's approach \cite{calcagno09} is based on Separation Logic over shapes (cf. sec.\ref{sect:SeparationLogic}).
The method approximates both sides of a rule to perform abductive reasoning.
This approach compares potentially matching rules, namely their beginnings, and in the case of multiple matching candidates, the most extended rule is chosen first.

\section{Pointer Rotation}
The old but not outdated theory of "Pointer Rotations" \cite{suzuki82} suggests pointer rotation and translations.
For example, given disjunctive pointers a barrel pointer rotations may be:
$$ \begin{pmatrix}
  x_1 & x_2 & x_3 & \cdots x_{n-1} & x_n\\
  x_2 & x_3 & x_4 & \cdots x_n & x_1
 \end{pmatrix}$$

With this, $x_1$ points to the content of $x_2$.

A left translation (known as a \texttt{slide}) is congruent to a left barrel rotation.
Next, \texttt{x:=y} is equivalent to $slide(x,y)$.
The objective of any rotation or translation is to replace error-prone individual code with well-known stable operations.
In this case "\textit{safe}", operations over pointers act as some specification.
For example:

\begin{verbatim}
y=NULL;
while (x!=NULL) {
  temp = x.next;
  x.next = y;
  y = x;
  x = temp;
}
\end{verbatim}

is equivalent to the notation

\begin{verbatim}
y = NULL;
while (x!=NULL) rotL(x,x.next,y);
\end{verbatim}

Although it does not follow directly from Suzuki's paper, it can be noticed that \texttt{rot(x,x,y)} is identical to the mapping $id$.
In analogy to the previous rotation to the right, a left rotation $rotate()$ may be defined as:

\begin{verbatim}
y = NULL;
while (x!=NULL){
  temp = x;
  x = y;
  y = x.next;
  x.next = temp;
}
\end{verbatim}

Equality can now be introduced upon this rotation
$$rot_L^{-1}(x,x.next,y) \equiv rot_L(x.next,x,y).$$
So, algebraic properties may be defined and proven dependent on given translations and rotations.
So, one rotation may have different stereotypes depending on incoming parameters.
For instance, $rotate(x.next,x.next.next.y)$.
Here $y$ denotes a target list, where $x.next$ denotes the first argument, which will be moved.
$x.next.next$ is the second argument, a pointer, which is not modified.

This approach is of interest from a theoretical and a practical side.
However, the calculus is hardly applicable to little more complicated algorithms, such as trees. Hence more research is required.

A pointer rotation (including translation) is supposed to be "safe" by nature.
Nevertheless, that is far from true.
Let us consider $rotate(y,y.next)$.
This example deletes the first element in many configurations.
However, this may look different for $rotate(x,y,x.next)$, when (1) the heap's content does not change, (2) all elements remain valid before and after rotation, and (3) the amount of variables does not change.
Pointer rotation does not need garbage collection.
It also benefits from effective and safe list operations.
Although pointer rotation does not require further specification, a tiny modification in the calling parameters in rotations may already lead to hard to predict behaviour (often undesired behaviour, though).
It is observed very frequently when corner cases occur or on aliases, and the consequence is that behaviour changes totally and may even delete lists where it has never been thought possible at all.
Suzuki \cite{suzuki82} suggests "base rotations" to restrict to some safe subsets, but, unfortunately, without explicitly describing what "base" really denotes -- even if it is intuitively clear what is meant, but very hard to find a standard definition.
However, in practice, it is hard.
The motivation behind this is to compose "safe" rotations individually.

\section{Separation Logic}
\label{sect:SeparationLogic}
Separation Logic (SL) makes use of predicates to describe dynamic memory \cite{reynolds02}, \cite{berdine05-2}, \cite{reynolds09}, \cite{burstall72}.
The logic is defined over the spatial operation $\star$, a logical conjunction (operators $\wedge,\vee,\neg$).

\begin{center}
\begin{tabular}{lcl}
  $\Phi$ & $::=$ & $\underline{true} \ | \ \underline{false} \ | \ x \ | \ REL(f_j(\vec{x})) \ | \ Pred(f_j(\vec{x}))$\\
         &       & $\ | \ \  \neg \Phi \ | \ \Phi \star \Phi \ | \ \forall x.\Phi[x] \ | \ \exists x.\Phi[x]$
\end{tabular}
\end{center}

$REL(f_j(\vec{x}))$ includes the atomic heap $a\mapsto b$, where $b$ denotes a location, and $b$ denotes a valid value (e.g. a class object).
Location means either an identifier or an access expression over valid object fields.
It is later introduced by Parkinson \cite{parkinson05}.
However, the initial definition of SL does not allow defining objects.
$Pred()$ denotes a call to a previously defined predicate that may require parameters.

The spatial operator by Berdine and Reynolds implies a separability of heap components.
Strictly speaking, the $\star$-operator can not only divide a heap but can also be used to conjunct two heaps.
The former is undoubtedly intended, whereas the latter is not.
The latter significantly complicates analyses of a heap and its components.

Properties of the spatial operator after Reynolds \cite{reynolds02} and Berdine \cite{berdine05-2}:

\begin{tabular}[t]{ll}
 (1) & compactness:\\
     & $p \not \Rightarrow p \star p$, $p \star q \not \Rightarrow p$, if $\exists q,q \not \equiv \texttt{emp}$\\
 (2) & commutativity:\\
     & $p_1 \star p_2 \Leftrightarrow p_2 \star p_1$
\end{tabular}

\begin{tabular}[t]{ll}     
 (3) & associativity:\\
     & $(p_1 \star p_2) \star p_3 \Leftrightarrow p_1 \star (p_2 \star p_3)$\\
 (4) & neutral element:\\
     & $p \star \texttt{emp} \Leftrightarrow \texttt{emp}  \star p \Leftrightarrow p$
\end{tabular}

\begin{tabular}[t]{ll}
 (5) & distributivity:\\
     & $(p_1 \vee p_2) \star q \Leftrightarrow (p_1 \star q) \vee (p_2 \star q)$\\
     & $(p_1 \wedge p_2) \star q \Leftrightarrow (p_1 \star q) \wedge (p_2 \star q)$
\end{tabular}

\begin{tabular}[t]{ll}
(6) & quantification:\\
     & $(\exists x.p) \star q \Leftrightarrow \exists x.(p \star q)$, if $x \not \in FV(q)$\\
     & $(\forall x.p) \star q \Leftrightarrow \forall x.(p \star q)$, if $x \not \in FV(q)$
\end{tabular}\\\\

Predicate \texttt{emp} holds for a given empty heap.
Set $FV$ denotes all variables not bound for a given assertion.
A binary tree is considered as a parameterised predicate which can inductively be defined as follows:
$$btree(l)::=\texttt{nil} \ | \ \exists x.\exists y:\ l \mapsto x,y \star btree(x) \star btree(y)$$
Here, a binary tree is described by predicate $btree$ and some untyped symbol $l$ or is just empty, or $l$ points to a simply-linked list whose content is $x$ followed by non-overlapping $y$.

SL is Hoare-based (cf. sec.\ref{sect:HoareCalculi}) and substructural (cf. \cite{restall00}).
The latter implies constants are replaced, e.g. by boolean values.
Higher constants, e.g. $\underline{true}$, are partial.
They consume some heap and return a boolean meaning depending on the passed heap.
SL also uses symbols for structural meanings as constants.
In SL, according to the non-repetition principle structural rules consist of thinning (THIN) \cite{restall00}, substitution, and heap cells constants.
In specified rules, "," is replaced by "$\star$", which separates two non-intersecting and unique heaps, except when told otherwise.
Heaps are defined inductively.
Rules (PERMUTE) and (CUT) are subtractive.
In SL, thinning does not hold.
Hence heap may not repeat.
This property is handy when a heap may occur at most once.

\begin{center}
\begin{tabular}{c}
 \inference[THIN]{X,Y \vdash Z}{X,A,Y \vdash Z}\\\\
 \inference[CONTR]{X,A,A,Y \vdash Z}{X,A,Y \vdash Z}\\\\
 \inference[PERMUTE]{X,A,B,Y \vdash Z}{X,B,A,Y \vdash Z}\\\\
 \inference[CUT]{X \vdash A \quad U,A,Y \vdash Z}{U,X,Y \vdash Z}
\end{tabular}\\
\end{center}

The frame rule defines as:

\begin{center}
\begin{tabular}{c}
\inference{\{P\}C\{Q\}}{\{P\star F\}C\{Q\star F\}}
\end{tabular}
\end{center}

It means that when a subprocedure call $C$ does not change a heap component, namely it includes frame $F$, then the antecedent is sufficient to prove in order to show the Hoare triple holds without $F$.

Berdine \cite{berdine05-2} suggest SL-based unbound arithmetic over pointers with compositions including dynamically growing arrays and recursive procedures.
It attempts to define dynamic memory recursively over a closed set of built-in rules only.
Berdine et al. also ask whether heap verification is not a typing problem.
From that paper follows, e.g. the undecidability of unbound pointers leads to flaky garbage collection events even for elementary expressions for memory offsets and to a very coarse rule selection due to greedy heuristics.

Bornat \cite{bornat00} proposes a model very close to SL called "\textit{remote separation}".
The model takes objects into arrays (cf. sec.\ref{sect:HeapIntoStack}).
So, any object field turns into an individual pointer with a different convention on labelling and uniqueness.
For a lifted definition, he postulates that first-order predicates are sufficient.
The main achievement of Hurlin \cite{hurlin09} is the new access design pattern towards heaps for multi-threaded programs.
If a given heap is atomic, then it is undividable, and therefore it must be a trivial and valid case.
If a heap is not provable, then compulsory simplifications are not applied.

Parkinson \cite{parkinson05-2} represents an attempt at an object-oriented extension of the classic SL \cite{reynolds02} based on Java as the input programming language.
Modularity and inheritance are modelled within "\textit{inverting calling control}" and "\textit{Abstract Predicate Family}".
Bornat's access model \cite{bornat00} can be applied because the continuity property holds for frames over objects.
Bornat notices that dependencies between predicates define an order for predicate calls.
The paper implies that the predicates within allow the full definition of heap and stack, but it does not allow, for instance, to define arbitrary first-order predicates.
Assertion predicates, which syntactically and semantically strongly differ from predicates for the input language, do not restrict themselves w.r.t. types.
However, the use of symbols imposes several limitations, due to the non-symbolic implementations of symbolic variables.
Predicates use them as locals as known from imperative programming languages.
Parkinson proposes method calls from parent classes, static fields, object introspection, inner classes and quantified predicates for future research.

\section{Future Research}
Apart from already mentioned tendencies, research gaps, and recent debates, the author of this paper would like to stress the following future research propositions:

\begin{enumerate}
 \item Both models, as discussed earlier, do not recognise varying objects in structures.
 So, if the contents of both are identical, then by definition, those two cells must be identical.
 However, in practice, this hardly can be mandatory since a later copy must not touch another copy.
 \item Predicates that are excluded by built-in tactics are manually folded and unfolded.
 Hutton suggests a generative method called "Fold and Unfold" on functionals in the context of functional programming languages.
 Here, further research is suggested to automate proofs, especially since proof generation seems to be a complex problem.
 Those problems lead to predicates being manually configured and triggered whilst proving, e.g. by proof hints.
 \item The use of "hot" code --- code that can be altered during execution is unlikely to be considered.
 Some previously mentioned authors consider "unique possibilities on expressibility".
 However, this euphoric opinion should be rethought by all means since the practical possibilities gained are too limited.
 On one side, arbitrary code is inserted into a running system whose specification only may be available.
 On the other side, this introduces other security risks and especially down-grades performance for any application considerably.
 Expressibility is not extended.
 It is just the moment that code is loaded.
 Loaded code is presumably a discipline on code linkers, which implement security mechanisms as included in the "\textit{GCC}" and "\textit{binutils}" framework, rather than verifying code.
 Virtual machines and interpreters implement similar security mechanisms called "\textit{Binding programming interfaces}".
 \item The level of intuition and proof explanation are essential criteria for verifier acceptance.
 This number plays a significant role and generalised counter-example generations, which can be considered absent at the moment.
 \item Due to upcoming models and approaches, the pressure for integration raises.
 There is a need for interoperating tools on the different stages of verification.
 The main reason for this is its lack of adaption and extension of the intermediate representation.
 \item Too often, solely scenarios are considered for heap verifiers.
 Although heap verifiers by design are constructed to suffice specific scenarios, the lack of integration and overall revision diverge the gained benefit severely and only allow estimates of real problems solved.
 A standard code would be required to make accurate estimates, but a joint heap model base would also be needed.
 \item Once integration starts, more and more unification attempts are expected.
 \item Further parallelisation of Hoare calculi for heap verification is despite similar attempts in garbage collection are not expected since the research focus currently is more on fundamentals than on a clear direction of parallelisation.
 Parallelisation, however, is expected in bordering disciplines, especially in model checking techniques, since their parallel formula constraints may efficiently be checked.
 \item Expressibility and completeness issues remain open.
 Variable modes remain an active research area at least for the next 12 years for static, global and dynamic memory.
 \item Dodds \cite{dodds08} proposes a descriptive transformation language for dynamic memory.
 His approach strongly differs and does not seem applicable to imperative programming languages at first glance.
 Here, broader research is needed for C-dialects on applicability since synergy effects could be taken from Dodds transformation language.
 \item It is expected that the descriptive paradigm trends from the 2010s remain.
 However, the focus may change from a functional to a logical paradigm.
 Due to its advantages in favour of functionals, more and more reuse can be observed in recent verifier implementations, although currently only loose features yet.
 \item It is expected that algebraic theories will more and more advance logical rules due to the fastly growing complexity of programming language features.
 So, a higher acceptance barrier is expected to introduce new features rather than individual built-in functions.
 Notably, the analysis of logical expressions has to reduce exponential complexity ideally to linear.
 \item In analogy to the previous discussion on dynamic code loading, higher-order Hoare calculi may not play the role estimated by its proponents.
 \item Partial specification as specification simplification technique is expected to raise meaning due to its pragmatism.
\end{enumerate}

%%%%%%%%%%%%%%%%%%%%%%%%%%%%%%%%%%%%%%%%

% REFERENCES:
%\printbibliography

\end{document}